# Electrical Image Potential and Solvation Energies for an Ion in a Pore in a Metallic Electrode or in a Nanotube

J. B. Sokoloff, Physics Department, Northeastern University, Boston, MA and Physics Department, Florida Atlantic University, Boca Raton, FL

**E**lectrical image potentials can be important in small spaces, such as nanoscale pores in porous electrodes, which are used in capacitive desalination and in supercapacitors, as argued by Bazant's group at MIT. It will be shown here that inside pores in porous metallic materials the image potentials can be considerably larger than near flat walls, as a result of the fact that the dielectric constant for an electric field perpendicular to a wall is much smaller than the bulk dielectric constant of water. Calculations will be presented for the image potential in spherical and cylindrically shaped pores. The calculations for cylindrical pores can also be applied to nanotubes. It was believed for a long time, on the basis of molecular dynamics simulations, that in order to push a salt solution through a small radius nanotube, work must be done against the solvation energy of the ions, which is larger inside a narrow nanotube than it is in the bulk. The relatively large image charge potential energy in narrow nanotubes, however, tends to oppose this increase in the solvation energy. The degree to which the image potential facilitates the flow of the salt ions into nanotubes will be discussed.

## I. Introduction

Porous electrodes play an essential role in capacitive desalination[1-8] and supercapacitors[9-13]. In order to determine the ability of a porous material to absorb ions from a solution, it is essential to understand the various contributions to the energy of the ions in nanometer scale pores. It was pointed out in Ref. 4 that the electrical image potential energy (i.e., the interaction of an ion with charge that it induces in the walls of a pore) makes an important contribution to the energy of the ions in the pores. The electrical image potential energy also makes an important contribution to the energy of ions in nanotubes. In particular, they are able to make up for the loss of solvation energy when an ionic solution enters a nanotube, possibly making it possible for ions to flow into even relatively narrow nanotubes. Here, the electrical image potential and solvation energy of ions within a spherically shaped and cylindrically shaped pore will be studied.

## II. Spherical Pores

First, let us consider a spherical shell of radius *a* with a uniform charge *q* at the center of a spherical pore of radius *b* in a metallic electrode. The image is a spherical shell of radius *R*, as illustrated in Fig. 1.

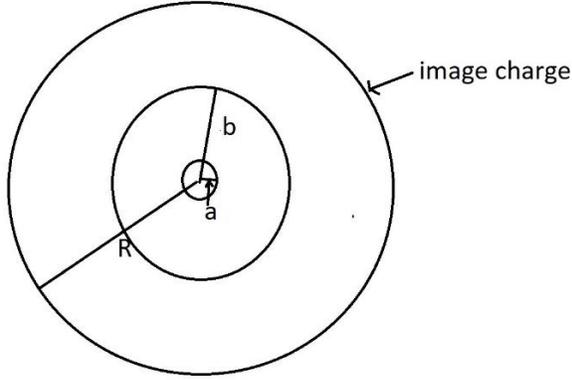

Fig. 1: This illustrates a charged ion of radius *a* at the center of a spherical cavity of radius *b*, which results in an image charge, which is a spherical shell of radius *R*.

The permittivity inside the pore is likely a tensor, which when diagonalized has an r-r diagonal element equal to $\varepsilon_{perp}$ which is considerably smaller than the permittivity in bulk water. For the case of an ion at the center of the pore, the electrical potential *V* satisfies Poisson's equation

$$\varepsilon_{perp} \frac{1}{r^2} \frac{d}{dr}\left(r^2 \frac{dV}{dr}\right) = 0. \qquad (1)$$

The image charge, equal to $-q(a/r)$, is distributed uniformly over the image sphere. Then the total potential at a distance r from the center of the pore is given by

$$V = \frac{q}{4\pi\varepsilon_{perp} r} - \frac{b}{r}\frac{q}{4\pi\varepsilon_{perp}(b^2/r)} = \frac{q}{4\pi\varepsilon_{perp} r} - \frac{q}{4\pi\varepsilon_{perp} b}. \qquad (2)$$

Consequently, the electrical image potential energy of the ion is given by

$$U_{image} = -\frac{q^2}{4\pi\varepsilon_{perp} b} = -\frac{2\varepsilon_\parallel}{\varepsilon_{perp}}\frac{a}{b}\frac{q^2}{8\pi\varepsilon_\parallel a}, \qquad (3)$$

where $\varepsilon_\parallel$ is the permittivity parallel to the pore's wall, since the second term in Eq. (2) is the potential due to the charge induced in the walls of the pore. Its interaction with the charge *q* is the image potential energy. Since $\varepsilon_{perp}$ is $2.1\varepsilon_0$ [14] (assuming that $\varepsilon_{perp}$ for a curved surface is comparable to its value for a flat surface), where $\varepsilon_0$ is the permittivity of free space, whereas the permittivity assumed in Ref. 4 was between $8\varepsilon_0$ and $16\varepsilon_0$, the image potential energy for an ion at the center of a pore with *b* of the order of a nanometer is larger than the estimate in Ref. 4.

The image potential which lowers the energy of an ion when it is inside the pore is opposed by the self-energy, which is larger inside the pore than in the bulk fluid. In order to calculate the self-energy at the center of the pore, we use the electric field due to a charged spherical shell of radius *a* at the center of the pore, as a model for the ion, without the contribution from the charge rearrangement on the boundary of the pore[16], It is given by

$$E = \frac{q}{4\pi\varepsilon_{perp} r}. \quad (4)$$

The Born self-energy[15] is then given by

$$U = (1/2)\varepsilon_{perp}(4\pi)\int_a^\infty r^2 dr E^2 = \frac{q^2}{8\pi\varepsilon_{perp} a} = \frac{\varepsilon_\|}{\varepsilon_{perp}} \frac{q^2}{8\pi\varepsilon_\| a}. \quad (5)$$

If $\varepsilon_\| = \varepsilon = 81\varepsilon_0$, $\varepsilon_{perp} = 2.1\varepsilon_0$ [14], where $\varepsilon_\|$ is the permittivity component parallel to the wall of the pore, assumed to be comparable to the permittivity of the bulk fluid $\varepsilon$, we find from Eqs. (3) and (5) that $U_{image} < U$ at the center of the pore by a factor of $a/b$, $U$ inside the pore is a factor $\varepsilon_\|/\varepsilon_{perp} = 38.6$ larger than the Born self-energy in the bulk fluid, which is equal to $q^2/(8\pi\varepsilon a)$. [Eq. (5) can also be obtained by calculating the interaction of the charge on the surface of the spherical shell with itself, as was done in the second section of Ref. 15.] In the case of capacitive desalination, the energy barrier for an ion to enter the pore due to $U$ can be overcome by the voltage difference between the bulk solution and the electrode if it exceeds the difference between $U$ in the pore and $U$ in the bulk solution divided by $q$, which is equal to *2V*. Ionic screening of the electric field, however, can provide a sizeable reduction in the self-energy.

Using Nordblom's screening theory[15,17], which is valid for high ion concentrations, we have

$$E = \frac{Q}{4\pi\varepsilon_{perp} r^2}\left(1 - \frac{r^3 - a^3}{h^3 - a^3}\right)\theta(r-a)\theta(h-r), \quad (6)$$

and hence,

$$U = (1/2)\varepsilon_{perp}(4\pi)\int_a^\infty r^2 dr E^2 = \frac{q^2}{8\pi\varepsilon_{perp}(h^3-a^3)^2}\left(\frac{h^6}{a} + h^3 a^2 - \frac{9h^5 + a^5}{5}\right)$$
$$= \frac{q^2}{8\pi\varepsilon_{perp} a}\Gamma, \quad (7)$$

where

$$\Gamma = \left(\frac{h^3}{a^3} - 1\right)^{-2}\left(\frac{h^6}{a^6} + \frac{h^3}{a^3} - 1.8\frac{h^5}{a^5} - 0.2\right). \quad (8)$$

Some values of $\Gamma$ are given in table I. The correlation hole approach to screening used in Ref. 4, appears to be equivalent to the approach used in Ref. 17 in the high ion density limit. The "correlation hole" referred to in Ref. 4 is identical to the screening charge sphere of radius *h* discussed above when *a<<h*, but the maximum ion concentration in the data fit with the model presented in Ref. 4 of 60mM ($3.61\times10^{25} m^{-3}$) is considerably smaller than the concentration at which one expects the Nordblom theory[17] is expected to be valid. On the basis of the values of $\Gamma$ given in the table, the screening will not be sufficient to reduce the self-energy at the center of the pore by a sufficient amount to make it smaller than the self-energy in the bulk fluid. This implies that the ions would not be able to enter the pore, in the absence of an electrical potential difference due to an external source between the bulk solution and the pore.

**Table I:** $\Gamma$ is given for several values of $n_B$, including $3.65 \times 10^{26} m^{-3}$, the salt concentration of sea water and $3.65 \times 10^{27} m^{-3}$, the salt concentration at the solubility limit of sodium chloride.

| $n_B$ | $h/a$ for an $Na^+$ ion | $\Gamma$ for an $Na^+$ ion | $h/a$ for a $Cl^-$ ion | $\Gamma$ for a $Cl^-$ ion |
|---|---|---|---|---|
| $10^{26} m^{-3}$ | 11.5 | 0.848 | 8.00 | 0.793 |
| $3.65 \times 10^{26} m^{-3}$ | 7.48 | 0.771 | 5.20 | 0.672 |
| $2 \times 10^{27} m^{-3}$ | 4.24 | 0.594 | 2.99 | 0.475 |
| $3 \times 10^{27} m^{-3}$ | 3.73 | 0.560 | 2.63 | 0.411 |
| $3.65 \times 10^{27} m^{-3}$ | 3.50 | 0.517 | 2.47 | 0.383 |

If the center of the ion were within a distance *a* from the wall of the pore which is much less than *b* (assuming that *a*<<*b*), the ion would "see" the surface of the pore as a plane to a good approximation, and hence, from table III in Ref. 15 for the image potential resulting from a metallic planar surface, we obtain for an ion located at the wall

$$U_{image} \approx -1.3 \frac{q^2}{8\pi\varepsilon_\parallel (b-d)} = -1.3 \frac{q^2}{8\pi\varepsilon_\parallel a}, \qquad (9)$$

where *d* is the distance of the center of the ion from the center of the pore, for $\varepsilon_\parallel = 81\varepsilon_0$, $\varepsilon_{perp} = 2.1\varepsilon_0$. The image potential as a function of the distance from a flat wall is plotted in Fig. 2 as a function of its distance *z* from the wall below using Eq. (26) in Ref. 15. Also included is a plot of $U_{image}(z=a)(a/z)$ (the lower plot). This shows that $U_{image}(z)$ is approximately inversely proportional to *z*.

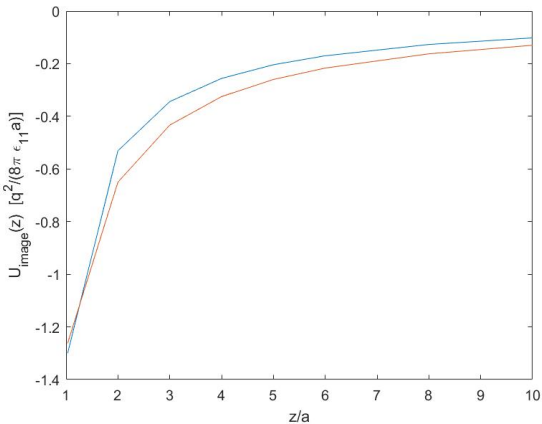

Fig. 2: This is a plot of the image potential energy for a charged sphere of radius *a* as a function of the distance *z* of its center from a plane metallic wall.

Also, the self-energy near a plane wall (and hence near the wall of the spherical pore) from table I in Ref. 15 is

$$U \approx 2.55 \frac{q^2}{8\pi\varepsilon_\parallel a}. \tag{10}$$

An approximate calculation of the image potential for a point ion located away from the center of the pore, in the limit of small $\varepsilon_{perp}/\varepsilon_\parallel$, on the basis of a solution of Poisson's equation for the tensor permittivity as an expansion in d/b, where d is the distance of the ion from the center of the pore, is given in appendix A. The resulting image potential from Eqs. (A25,A26) is given by

$$U_{image} = -\frac{q^2}{4\pi\varepsilon_{perp}b}\left[1 + A\left(\frac{\varepsilon_{perp}}{\varepsilon_\parallel}\right)^{1/2}\right], \tag{11}$$

where

$$A = \frac{b}{4\pi d}\sum_{\ell=1}^{\infty}[\ell(\ell+1)]^{1/2} v^{[\ell(\ell+1)]^{1/2}}, \tag{12}$$

where $v = (d/b)^{(\varepsilon_\parallel/\varepsilon_{perp})^{1/2}}$, which is given in table II below.

**Table II:** This table gives the parameter $A$ as a function of d/b.

| d/b | A |
| --- | --- |
| 0.2 | $4.08 \times 10^{-7}$ |
| 0.7 | 0.0084 |
| 0.8 | 0.0321 |
| 0.9 | 0.180 |
| 0.95 | 0.792 |

The summation for $A$ in Eq. (12) diverges as d/b approaches 1, indicating that this solution most likely breaks down as d/b approaches 1. Since for relative small values of d/b, A is smaller than 1, it appears that for ions that are far from the walls of the pore, $U_{image}$ is of the order of the value given in Eq. (3). The image potential at the center of the pore, given by Eq. (3) is larger than its value near the wall if

$$\frac{2\varepsilon_\parallel}{\varepsilon_{perp}}\frac{a}{b} > 1.3. \tag{13}$$

On the basis of Eq. (13) we find that for $\varepsilon_\parallel = 81\varepsilon_0$, $\varepsilon_{perp} = 2.1\varepsilon_0$, if a/b>0.017, the image potential at the center of a spherical pore is larger than it would be near a flat wall or between two parallel walls or near the wall of a spherical pore. The value of $U + U_{image}$ near the wall of the pore can be less than the value of $U$ in the bulk fluid for the screening that occurs for an ion concentration of $2 \times 10^{27} m^{-3}$ or more[15]. This is possible because if the ion is at the wall and a<h, the screening due to the other ions does not eliminate the image potential energy of the ion. From Eqs. (3) and (5), however, it appears that $U + U_{image}$ inside and away from the walls of the pore will be larger than $U$ in the bulk fluid. The treatment of nanopores in Ref.18 considers the Born self-energy but does not include the tensor nature of the permittivity near a surface and the electrical image potential energy.

### III. Cylindrical Pores and Nanotubes

Let us now consider the image potential energy for a point charge to represent an ion a distance $d$ from the axis of a metallic cylinder of radius $b$. The Coulomb potential in cylindrical coordinates for an ion lying on the cylinder's axis is the solution to Poisson's equation

$$\varepsilon_{perp}\left(\frac{\partial^2 V}{\partial \rho^2}+\frac{1}{\rho}\frac{\partial V}{\partial \rho}\right)+\varepsilon_{\|}\left(\frac{\partial^2 V}{\partial z^2}+\frac{1}{\rho^2}\frac{\partial^2 V}{\partial \phi^2}\right)=-\frac{q}{\rho}\delta(\rho-d)\delta(\phi)\delta(z). \tag{14}$$

The details of the solution of Eq. (14) are given in Appendix B by adapting the solution of Poisson's equation in Ref. 18. We obtain for the image potential from Eq. (B26)

$$U_{image}=-\frac{q^2}{8\pi\varepsilon_{perp}b}\frac{4}{\pi}\sum_{m=0}^{\infty}\int_0^{\infty}dk'\frac{I_{mu^{1/2}}(k'u^{1/2}d/b)^2 K_{mu^{1/2}}(k'u^{1/2})}{I_{mu^{1/2}}(k'u^{1/2})}, \tag{15}$$

where $I_a(x), K_a(x)$ are the modified Bessel functions of order $a$ and where $k'=kb$. In table III, the results for $U_{image}$ are shown for $\varepsilon_{\|}=81$, $\varepsilon_{perp}=2.1$, where $I'=U_{image}(8\pi\varepsilon_{perp}b/q^2)$.

**Table III:** This table gives $I'$ as a function of d/b.

| d/b   | I'    |
|-------|-------|
| 0.    | 0.280 |
| 0.2   | 0.283 |
| 0.3   | 0.287 |
| 0.5   | 0.300 |
| 0.7   | 0.329 |
| 0.9   | 0.446 |
| 0.95  | 0.599 |

Since Ref. 19 shows that for tubes with radii of the order of a nanometer, $\varepsilon_{\|}$ becomes larger than the bulk water permittivity of $81\varepsilon_0$, let us calculate $U_{image}$ for larger values of $\varepsilon_{\|}$. For example, for d=0 and for $\varepsilon_{\|}=200\varepsilon_0$, $U_{image}=0.179\,(q^2/8\pi\varepsilon_{perp}b)$ and for $\varepsilon_{\|}=300\varepsilon_0$, $U_{image}=0.146\,(q^2/8\pi\varepsilon_{perp}b)$.

If we apply the results given in table III to pores in porous electrodes, we conclude that for a pore with a radius of the order of a nanometer, the image potential is noticeably larger than the value used in Ref. 4. It is difficult to evaluate $U_{image}$ for $d$ close to $b$, but since the wall should appear to the ion as being nearly flat, it should be well approximated by Eq. (9) [15], for b-d=a, for an ion of radius $a$ in contact with the wall. Since inside the cylinder

$$U_{image} = -I' \frac{\varepsilon_{\|}}{\varepsilon_{perp}} \frac{a}{b} \frac{q^2}{8\pi \varepsilon_{\|} a}, \qquad (16)$$

The self-energy can be calculated for an ion at the center of the cylinder as follows: Since from Eqs. (B19-B22) in the appendix the Green's function is given by

$$G(\vec{r}-\vec{r}') = \sum_{m=-\infty}^{\infty} e^{im(\phi-\phi')} \int_0^{\infty} dk J_{u^{1/2}m}(u^{1/2}k\rho) J_{u^{1/2}m}(u^{1/2}k\rho') e^{-k(z_>-z_<)}, \qquad (17)$$

with $u = \varepsilon_{\|}/\varepsilon_{perp}$, where for calculating the self-energy of a charged spherical shell model for the ion, we will set $\rho = (a^2-z^2)^{1/2}$, $\rho' = (a^2-z'^2)^{1/2}$ and $z_>$ is the larger of $z$ and $z'$ and $z_<$ is the smaller of $z$ and $z'$. This is a Green's function because $e^{-ik(z_>-z_<)}$ is the Green's function for the operator $d^2/dz^2 - k^2$. It is the solution of the equation

$$\left( \frac{\partial^2 G(\vec{r}-\vec{r}')}{\partial \rho^2} + \frac{1}{\rho} \frac{\partial G(\vec{r}-\vec{r}')}{\partial \rho} \right) + u \left( \frac{\partial^2 G(\vec{r}-\vec{r}')}{\partial z^2} + \frac{1}{\rho^2} \frac{\partial^2 G(\vec{r}-\vec{r}')}{\partial \phi^2} \right) = -\frac{1}{\varepsilon_{perp} \rho} \delta(\rho-d) \delta(\phi) \delta(z).$$
(18)

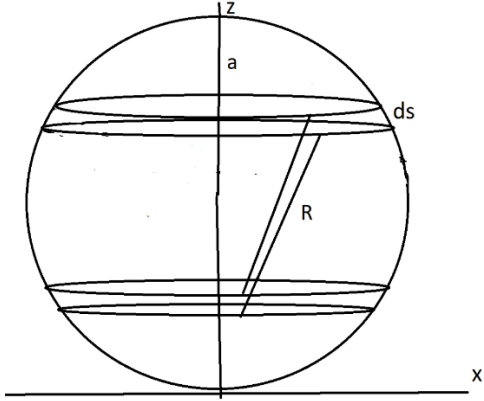

Fig. 3: Illustration of the calculation of the self-energy of an ion modeled as a charged spherical shell of radius $a$.

Then $U$ is given by

$$U = \frac{1}{2} \frac{(\sigma a)^2}{4\pi \varepsilon_{perp}} \iint \Big|_{|\vec{r}|,|\vec{r}'|=a} d^3r d^3r' G(\vec{r}-\vec{r}')$$

$$= \frac{1}{2} \frac{(\sigma a)^2}{4\pi \varepsilon_{perp}} \int_0^{2\pi} d\phi' \int_0^{2\pi} d\phi \sum_{m=-\infty}^{\infty} e^{im(\phi-\phi')} \int_{-a}^{a} dz' \int_{-a}^{a} dz \int_0^{\infty} dk J_{u^{1/2}m}(u^{1/2}k\rho) J_{u^{1/2}m}(u^{1/2}k\rho') e^{-k(z_>-z_<)}$$
, (19)

as illustrated in Fig. 3, where $\rho = (a^2-z^2)^{1/2}/a$, $\rho' = (a^2-z'^2)^{1/2}/a$, since the element of area is given by $dA = a \sin\theta d\theta a d\phi = a^2 dz d\phi$ (where $\theta$ is the azimuthal angle from spherical coordinates), and similarly for the primed quantities. Doing the integrals over $\phi$ and $\phi'$ and making the substitution $k' = u^{1/2}k$, we obtain

$$U = \frac{q^2}{32\pi u^{1/2}\varepsilon_{perp}a^2}\int_{-a}^{a}dz'\int_{-a}^{a}dz\int_{0}^{\infty}dk' J_0(k'\rho)J_0(k'\rho')e^{-k'u^{-1/2}(z_>-z_<)}. \tag{20}$$

Doing the integral with $u=81/2.1$, gives

$$U = \frac{q^2}{32\pi u^{1/2}\varepsilon_{\|}a^2}u\int_{-a}^{a}dz'\int_{-a}^{a}dz\int_{0}^{\infty}dk' J_0(k'\rho)J_0(k'\rho')e^{-k'u^{-1/2}(z_>-z_<)} = \frac{u^{1/2}I}{4}\frac{q^2}{8\pi\varepsilon_{\|}a} = 8.86\frac{q^2}{8\pi\varepsilon_{\|}a}, \tag{21}$$

where $I$ is the above triple integral in Eq. (20). The numerical value given for the quantity $u^{1/2}I/4$ is for $u = \varepsilon_{\|}/\varepsilon_{perp} = 81/2.1$. $U$ seems to be independent of $b$. This likely occurs because $U$ only depends on the form of the tensor permittivity, which has been assumed to have the form that was used to calculate $U$ above no matter how large we choose $b$ to be. In reality, however, if the radius is large enough, the permittivity at the center of the tube will revert to its bulk water value. If $\varepsilon_{\|} = \varepsilon_{perp} = \varepsilon$, $I=4$ and hence $U$ reduces to the Born value for a scalar permittivity of $U = q^2/(8\pi\varepsilon a)$. (Incidentally, the calculations of the of the self-energy and the image potential energy performed in Ref. 15 by an approximation valid to lowest order in $\varepsilon_{perp}/\varepsilon_{\|}$ can also be done without any approximation by direct integration. Identical numerical results are obtained.)

Let us now give some numerical illustrations of the effects of $U_{image}$ and $U$ for a solution of sodium chloride. Near the wall (a distance $d-a<<b$), $U_{image}$ is given by Eq. (9) and $U$ is given by Eq. (10). We found that at the center of the tube $U$ is given by Eq. (21). If $a=0.167nm$ (i.e., for chlorine ions) and $b=1nm$, $U_{image}$ at the center of the cylinder given by Eq. (16) is

$$U_{image} = -1.80\frac{q^2}{8\pi\varepsilon_{\|}a}, \tag{22}$$

without considering the effect of screening on the image potential. (Since the screening length $h$ is smaller than the distance of the ion from the cylinder's wall, because of screening, the image potential energy is not likely to have a significant effect for ions located a distance greater than $h$ from the cylinder's wall.) Hence, even if the ionic concentration in the pore is close to the solubility limit (i.e., the ion concentration $3.65\times10^{27}m^{-3}$), which gives, using the screening factor $\Gamma$ given in table I for this concentration, the screened self-energy given by

$$U = 3.39\frac{q^2}{8\pi\varepsilon_{\|}a}, \tag{23}$$

from Eq. (7,21) and table I. Therefore, $U+U_{image}$ within the cylinder will still be greater than $U$ in the bulk solution, which is equal to

$$U = \frac{q^2}{8\pi\varepsilon_\parallel a}. \qquad (24)$$

This will certainly be the case unless the ion is within a distance from the wall less than $h$. Sufficiently near the walls, however, for an ion solution with concentration $3\times10^{27} m^{-3}$, $U + U_{image}$ will be less than the value of $U$ in the bulk solution[15]. In the case of capacitive desalination, the solvation energy can be overcome by the external potential applied to the electrodes if the potential difference between the bulk solution and inside the pore or nanotube is greater than the difference between Eqs. (23) and (24) divided by $q$, which is equal to *0.127V*. If b were equal to 0.4 nm, however, $U_{image} = -5.07 q^2/(8\pi\varepsilon_\parallel a)$, for chloride ions, from table III and Eq. (16), and hence $U + U_{image}$ at a distance less than *d=h=0.269nm* from the walls of the pore or nanotube (corresponding to $d/b \approx 0.7$), using the value of $U$ given by Eq. (23), will be less than the value of $U$ in the bulk solution. Although it is known that the continuum approximation used in this article is no longer valid (as the water forms one dimensional chains)[21-28], this result suggests the importance of the image potential energy in allowing ions to occupy nanotubes or nanometer pores in electrodes.

Perhaps a more realistic, although crude, picture for an ion in a nanotube that is sufficiently narrow so that water enters the tube as a one dimensional chain is the following: We assume that inside the tube, the permittivity is equal to $\varepsilon_0$ and the ion only interacts with the dipole moment of nearest neighbor water molecules with an interaction of the order of

$$U_e = 2\frac{1}{4\pi\varepsilon_0}\frac{qp}{r^2}, \qquad (25)$$

where $p = 6.17\times 10^{-30} C\cdot m$ is the dipole moment of a water molecule and $r$ is the distance between the center of the ion and the center of the one of the nearest neighbor water molecules. If we take it to be equal to $3\times 10^{-10} m$, we obtain $U_e = -1.23 eV$. Then, the solvation energy $\Delta U$, which is the difference between the self-energy of the ion inside the tube and in the bulk solution, is

$$\Delta U = \frac{q^2}{8\pi\varepsilon_0 a}\left(1 - \frac{\varepsilon_0}{\varepsilon}\right) + U_e = 3.03 eV, \qquad (26)$$

substituting $a = 1.67\times 10^{-10} m$ and $\varepsilon = 81\varepsilon_0$. The image potential energy for an ion at the center of the tube found from Eq. (15) with $\varepsilon_\parallel = \varepsilon_{perp} = \varepsilon_0$ is given by

$$U_{image} = -1.74\frac{q^2}{8\pi\varepsilon_0 a} = -3.13 eV, \qquad (27)$$

which means that

$$\Delta U + U_{image} = -0.10 eV, \qquad (28)$$

and hence, the total of the self-energy and the image potential inside the tube is lower than the self-energy outside of the tube.

So far, it has been assumed that the cylinder's wall is metallic. If the cylinder is taken to represent a single wall nanotube, it can be assumed to be a two dimensional conductor if the electron mean free path is smaller than the circumference. If not, it behaves as a 1-d conductor (with conduction only along the z-axis, the axis of the tube). Even if the tube behaves as a one dimensional conductor, it will still need to be an equipotential, because of the following argument: The z component of the electric field is given by

$$E_z = -\frac{\partial V}{\partial z} = \frac{q}{2\pi^2 \varepsilon_{perp}} \sum_{m=-\infty}^{\infty} e^{im\phi} \int_0^\infty dk\, k \sin kz \left[ I_{mu^{1/2}}(ku^{1/2}\rho_<) K_{mu^{1/2}}(ku^{1/2}\rho_>) + A_m(k) I_{mu^{1/2}}(ku^{1/2}\rho) \right]. \quad (29)$$

If the tube is a one-dimensional conductor, when $\rho_< = d$, $\rho_> = \rho = b$,

$$0 = E_z = -\frac{\partial V}{\partial z} = \frac{q}{2\pi^2 \varepsilon_{perp}} \sum_{m=-\infty}^{\infty} e^{im\phi} \int_0^\infty dk\, k \sin kz \left[ I_{mu^{1/2}}(ku^{1/2}d) K_{mu^{1/2}}(ku^{1/2}b) + A_m(k) I_{mu^{1/2}}(ku^{1/2}b) \right],$$

(30)

which gives the above value for $A_m(k)$. This means that

$$0 = E_\phi = -\frac{1}{\rho} \frac{\partial V}{\partial \phi}$$
$$= -\frac{q}{2\pi^2 \varepsilon_{perp}} i \sum_{m=-\infty}^{\infty} m e^{im\phi} \int_0^\infty dk \cos kz \left[ I_{mu^{1/2}}(ku^{1/2}\rho_<) K_{mu^{1/2}}(ku^{1/2}\rho_>) + A_m(k) I_{mu^{1/2}}(ku^{1/2}\rho) \right] \quad (31)$$

on the surface of the cylinder where $\rho_< = d$, $\rho_> = \rho = b$, implying that the wall is an equipotential. In Misra and Blankschtein's simulation[29], graphene is assumed to be an perfect insulator, and thus, there is no electrical image potential energy. As long as the walls have any nonzero conductivity, however, when the ion is stationary, there will always be an image charge potential energy, as long as the one waits long enough for electrons in the wall to flow towards or away from the region in the wall opposite the ion. If the ion is moving parallel to the wall, however, it was shown in appendix B of Ref. [30] that the image charge lags behind the ion by a distance parallel to the wall of

$$\Delta x \approx \frac{\varepsilon_\| v}{\sigma_{2d}}, \quad (32)$$

where $v$ is the velocity of the ion and $\sigma_{2d}$ is the electrical conductance of the wall. Thus, the ion will interact with the wall with an electrical image potential energy unless $\sigma_{2d}$ is sufficiently small so that $\sigma_{2d}/\varepsilon_\| \ll v$.

### IV. Treatment of the Interaction of an Ion with Insulating Walls

In Misra and Blankschtein's simulation[29], the interaction of an ion with the wall results from polarization of the individual carbon atoms, which was treated by a model in which the polarization of the graphene is assumed to result from displacement of the electrons on a carbon atom with respect to its nucleus. In order to make contact with the polarization interaction in Misra and Blankschtein's simulation[29], consider the following continuum model for the polarization interaction between a point ion of charge $q$ and a 2-d surface:

$$V_{pol}(\vec{r}) = \frac{1}{4\pi\varepsilon_0} \sum_j \frac{\vec{p}_j \cdot (\vec{r}-\vec{r}_j)}{|\vec{r}-\vec{r}_j|^3} \approx \frac{1}{4\pi\varepsilon_0} \int d^2 r' \frac{\vec{\sigma}_p(\vec{r}') \cdot (\vec{r}-\vec{r}')}{|\vec{r}-\vec{r}'|^3}, \quad (33)$$

where the integral is over the surface of the cylinder and where $V_{pol}(\vec{r})$ is the potential at $\vec{r}$ due to the dipole moment induced in the surface per unit area $\vec{\sigma}_p(\vec{r}')$, created by the electric field due to an ion located at the point $\vec{r}_0$. Then, $\vec{\sigma}_p(\vec{r}')$ is given by

$$\vec{\sigma}_p(\vec{r}') = \frac{1}{4\pi\varepsilon_0} \chi_p \frac{q(\vec{r}'-\vec{r}_0)}{|\vec{r}'-\vec{r}_0|^3}, \quad (34)$$

where $\chi_p$ is the susceptibility that gives the dipole moment per unit area induced by the ion's electric field. Then, substituting for $\vec{\sigma}_p(\vec{r}')$ in the above expression for $V_{pol}(\vec{r})$, evaluated at $\vec{r}=\vec{r}_0=z_0\hat{z}$ gives the polarization interaction between the ion and the surface:

$$U_{pol} = qV_{pol}(\vec{r}_0) = \frac{q^2 \chi_p}{(4\pi\varepsilon_0)^2} \int dx'dy' \frac{1}{(z_0^2+x^2+y^2)^2} = \frac{q^2 \chi_p}{16\pi\varepsilon_0^2 z_0^2}. \quad (35)$$

It is proportional to $z_0^{-2}$, in contrast to the electrical image potential energy for a metallic surface, which is proportional to $z_0^{-1}$. When there is water present,

$$V_{pol}(\vec{r}) = \frac{1}{4\pi\varepsilon_\parallel} \int d^2 r' \frac{\sigma_{pz}(z-z') + (\varepsilon_{perp}/\varepsilon_\parallel)[\sigma_{px}(x-x') + \sigma_{py}(y-y')]}{\left[(z-z')^2 + (\varepsilon_{perp}/\varepsilon_\parallel)[(x-x')^2+(y-y')^2]\right]^{3/2}} \quad (36)$$

and

$$\vec{\sigma}_p(\vec{r}') = \chi_p \frac{q}{4\pi\varepsilon_\parallel} \frac{(z'-z_0)\hat{z} + (\varepsilon_{perp}/\varepsilon_\parallel)(x'^2+y'^2)}{\left[(z'-z_0)^2 + (\varepsilon_{perp}/\varepsilon_\parallel)(x'^2+y'^2)\right]^{3/2}}. \quad (37)$$

Substituting in the expression for $V_{pol}(\vec{r})$ at $\vec{r}=\vec{r}_0=z_0\hat{z}$, we obtain

$$U_{pol} = qV_{pol}(\vec{r}_0) = \frac{\chi_p q^2}{(4\pi\varepsilon_\parallel)^2} 2\pi \int_0^\infty \frac{\rho' d\rho'}{[z_0^2+(\varepsilon_{perp}/\varepsilon_\parallel)\rho'^2]^2} \approx \frac{\chi_p q^2}{16\pi\varepsilon_\parallel \varepsilon_{perp} z_0^2}. \quad (38)$$

The susceptibility is given by

$$\chi_p = 4\pi\varepsilon_0 n\alpha, \quad (39)$$

where $n$ is the number of carbon atoms per unit area and $\alpha$ is the polarizability of a carbon atom in cgs units[29], giving

$$U_{pol} = \frac{\varepsilon_0 n\alpha q^2}{4\varepsilon_\parallel \varepsilon_{perp} z_0^2}. \qquad (40)$$

Since $\varepsilon_{perp} \approx \varepsilon_0$ and $n\alpha$ is of the order of or a little smaller than a unit cell dimension, the image potential energy in the polarization model is of the order of or a little smaller than the image potential energy for metallic surface. In other words it is reduced by a factor of $\varepsilon_0 / \varepsilon_\parallel$ from its vacuum value.

For an ion at the center of a nonmetallic cylinder, the electrical potential produced by the ion at the origin is given in cylindrical coordinates by

$$V(\rho, z) = \frac{q}{4\pi\varepsilon_{perp}[(\varepsilon_\parallel / \varepsilon_{perp})\rho^2 + z^2]^{1/2}}, \qquad (41)$$

and hence,

$$\vec{E}(\rho, z) = \frac{q}{4\pi\varepsilon_{perp}} \left[ \frac{\rho(\varepsilon_\parallel / \varepsilon_{perp})\hat{\rho} + z\hat{z}}{[(\varepsilon_\parallel / \varepsilon_{perp})\rho^2 + z^2]^{3/2}} \right]. \qquad (42)$$

Since $\vec{\sigma}_p(\vec{r}) = \chi_p \vec{E}(\vec{r})$ and

$$U_{pol} = 2\pi bq \int_{-\infty}^{\infty} dz \vec{\sigma}_p \cdot \vec{E}(\vec{r}), \qquad (43)$$

or

$$U_{pol} = \frac{q^2 \chi_p b}{8\pi\varepsilon_{perp}^2} \int_{-\infty}^{\infty} dz \frac{b^2(\varepsilon_\parallel / \varepsilon_{perp})^2 + z^2}{[(\varepsilon_\parallel / \varepsilon_{perp})b^2 + z^2]^3} = \frac{q^2 \chi_p}{8\pi\varepsilon_{perp}^2 b^2} \int_{-\infty}^{\infty} d\bar{z} \frac{(\varepsilon_\parallel / \varepsilon_{perp})^2 + \bar{z}^2}{[(\varepsilon_\parallel / \varepsilon_{perp}) + \bar{z}^2]^3}$$
$$= \frac{q^2 \chi_p}{8\pi\varepsilon_{perp}^2 b^2} \frac{[3(\varepsilon_\parallel / \varepsilon_{perp}) + 1)]\pi}{8(\varepsilon_\parallel / \varepsilon_{perp})^{3/2}}, \qquad (44)$$

where $\bar{z} = z/b$. For $\varepsilon_\parallel / \varepsilon_{perp} = 81/2.1$,

$$U_{image} = 0.191 \frac{q^2 \chi_p}{8\pi\varepsilon_{perp}^2 b^2} = 0.095 \frac{q^2 n\varepsilon_0 \alpha}{\varepsilon_{perp}^2 b^2}, \qquad (45)$$

using the above expression for $\chi_p$, which makes it larger than the interaction with a flat insulating surface by a factor of $0.38\varepsilon_\parallel / \varepsilon_{perp}$ or a factor of 14.7 for the above values of the permittivity components. From table III,

$$U_{image} \, 0.28 \frac{q^2}{8\pi\varepsilon_{perp} b} \qquad (46)$$

for a metallic tube with the above values of the permittivity components. The ratio of the image potential energy for an insulating wall to that for a metallic wall is $1.54(\varepsilon_0 / \varepsilon_{perp})(n\alpha / a)$.

## V. Conclusions

It has been shown that electrical image potential energy can play an important role in allowing ions in an ionic solution to be absorbed by nanometer scale pores in metallic electrodes, but it appears that it is only in pores of size less than a nanometer that the sum of the ion's solvation energy and its image potential energy within the pore is smaller than the ion's solvation energy in the bulk solution, which would allow the ions to reside in the pore, unless the potential difference between the bulk solution and the inside of the pore exceeds the difference between the solvation energy in the bulk solution and the sum of the solvation and image potential energy inside the pore, which is only 0.127V. For a subnanometer radius cylindrical pore or a nanotube, however, the sum of the electrical image potential energy and the Born self-energy of an ion is able to be lower than the self-energy of an ion in the bulk water, making it possible for ions to spontaneously enter the pore or nanotube. The Born expression for the self-energy is believed to be an over-estimate[31-34]. Therefore, these conclusions can only be used to predict trends. In fact, for sodium ions, the solvation energy can be as much as a factor of 57% smaller than the value given by the Born approximation[16,31]; for chloride ions, it is only reduced by a factor of 94%. This implies that that positive ions will be more likely to be able to enter nanotubes and pores, because $U + U_{image}$ inside the tube or pore is more likely to be smaller than $U$ outside the tube or pore.

**Acknowledgements:** I wish to thank Alex Noy for discussions that I had with him about his unpublished work, which inspired some of the calculations in this article.

## Appendix A: Image Potential Energy for a Spherical Pore

Consider Poisson's equation in spherical coordinates[19],

$$\varepsilon_{perp} \frac{1}{r}\frac{\partial^2}{\partial r^2}(rV) + \varepsilon_{\|} \frac{1}{r^2 \sin\theta}\frac{\partial}{\partial \theta}\left(\sin\theta \frac{\partial V}{\partial \theta}\right) + \frac{\varepsilon_{\|}}{r^2 \sin^2\theta}\frac{\partial^2 V}{\partial \phi^2}$$
$$= q\frac{1}{r^2}\delta(r-r')\delta(\cos\theta - \cos\theta')\delta(\phi - \phi'). \quad (A1)$$

If the ion is assumed to lie on the z-axis, Poisson's equation has azimuthal symmetry. When a trial solution of the form $V(r,\theta) = r^{-1}R(r)P(\cos\theta)Q(\phi)$ is substituted in the associated homogeneous equation (i.e., Laplace's equation), we obtain

$$\frac{\varepsilon_{perp}}{R(r)}\frac{d^2 R(r)}{dr^2} + \frac{\varepsilon_{\|}}{r^2 \sin\theta P(\cos\theta)}\frac{d}{d\theta}\left(\sin\theta \frac{dP(\cos\theta)}{d\theta}\right) + \frac{\varepsilon_{\|}}{r^2 \sin^2\theta Q(\phi)}\frac{d^2 Q(\phi)}{d\phi^2} = 0. \quad (A2)$$

Let us set

$$\frac{1}{\sin\theta P(\cos\theta)}\frac{d}{d\theta}\left(\sin\theta \frac{dP(\cos\theta)}{d\theta}\right) = -\ell(\ell+1) + \frac{m^2}{\sin^2\theta}, \quad (A3)$$

and

$$\frac{1}{Q(\phi)}\frac{d^2Q(\phi)}{d\phi^2} = -m^2 \qquad (A4)$$

with $0 < \ell < \infty, -\ell < m < \ell$, whose solutions are spherical harmonics $Y_\ell^m(\theta,\phi)$, we find that $R(r)$ is a solution to the equation

$$\frac{d^2 R(r)}{dr^2} = \frac{\varepsilon_\parallel}{\varepsilon_{perp}} \frac{\ell(\ell+1)}{r^2}. \qquad (A5)$$

For $\ell \neq 0$ and large values of $\varepsilon_\parallel / \varepsilon_{perp}$, we can look for a solution of the form $R(r) = e^{u(r)}$. Substituting in the above equation we obtain

$$\left[\frac{d^2 u}{dr^2} + \left(\frac{du}{dr}\right)^2\right] e^u = \left(\frac{\varepsilon_\parallel}{\varepsilon_{perp}}\right)\frac{\ell(\ell+1)}{r^2} e^u. \qquad (A6)$$

Neglecting $d^2u/dr^2$, gives solutions of the form

$$u = \pm\left(\frac{\varepsilon_\parallel}{\varepsilon_{perp}}\right)^{1/2} [\ell(\ell+1)]^{1/2} \ln\frac{r}{b}, \qquad (A7)$$

where *b* is a constant. Since

$$\frac{d^2 u}{dr^2} = \mp\left(\frac{\varepsilon_\parallel}{\varepsilon_{perp}}\right)^{1/2} \frac{[\ell(\ell+1)]^{1/2}}{r^2} \qquad (A8)$$

and

$$\left(\frac{du}{dr}\right)^2 = \frac{\varepsilon_\parallel}{\varepsilon_{perp}} \frac{\ell(\ell+1)}{r^2}, \qquad (A9)$$

we are justified in neglecting the second derivative if $(\varepsilon_\parallel / \varepsilon_{perp})^{1/2}$ is sufficiently large. This approximation becomes better as $\ell$ increases. Since in our case $(\varepsilon_\parallel / \varepsilon_{perp})^{1/2}$ is only equal to 6.21, this solution is only a crude approximation. For $\ell = 0$, the solution for *R(r)/r* has the form

$$\frac{R(r)}{r} = A + \frac{B}{r}, \qquad (A10)$$

where A and B are constants to be determined by boundary conditions. For $\ell > 0$,

$$\frac{R(r)}{r} = r^{-1}\left[A'\left(\frac{r}{b}\right)^{\alpha_\ell} + B'\left(\frac{b}{r}\right)^{\alpha_\ell}\right], \qquad (A11)$$

where *A'* and *B'* are constants and $\alpha_\ell = [(\varepsilon_\parallel / \varepsilon_{perp})\ell(\ell+1)]^{1/2}$.

Following the derivation of the Green's function in Ref.[19], we look for a Green's function of the form

$$G(\vec{r},\vec{r}') = \sum_{\ell=0}^{\infty} \sum_{m=-\ell}^{\ell} g_\ell(r,r') Y_\ell^m{}^*(\theta',\phi') Y_\ell^m(\theta,\phi). \quad (A12)$$

Substituting in Poisson''s equation above we obtain the following equation for $g_\ell(r,r')$:

$$\frac{1}{r}\frac{d^2}{dr^2}[rg_\ell(r,r')] - \left(\frac{\varepsilon_\parallel}{\varepsilon_{perp}}\right)\frac{\ell(\ell+1)}{r^2} g_\ell(r,r') = \frac{1}{\varepsilon_{perp} r^2}\delta(r-r'). \quad (A13)$$

Integrating this equation over r from $r'-\delta$ to $r'+\delta$, where $\delta \ll 1$, we obtain

$$\frac{d}{dr}[rg_\ell(r,r')]_{r=r'+\delta} - \frac{d}{dr}[rg_\ell(r,r')]_{r=r'-\delta} = \frac{1}{\varepsilon_{perp} r'}. \quad (A14)$$

Using the above approximate solution to Laplace's equation, we can write

$$g_\ell(r,r') = C\frac{1}{r_<}\left(\frac{r_<}{b}\right)^{\alpha_\ell} \frac{b}{r_>}\left[\left(\frac{r_>}{b}\right)^{\alpha_\ell} - \left(\frac{b}{r_>}\right)^{\alpha_\ell}\right], \quad (A15)$$

for $\ell > 0$, where $r_>$ ($r_<$) is the larger (smaller) of $r$ and $r'$. We have chosen $b$ to represent the radius of the pore, where the potential must vanish. From the above continuity condition, we can write

$$\frac{d}{dr}[rg_\ell(r,r')]_{r=r'+\delta} - \frac{d}{dr}[rg_\ell(r,r')]_{r=r'-\delta} =$$
$$\frac{C}{r'}\left(\frac{r'}{b}\right)^{\alpha_\ell} \alpha_\ell\left[\left(\frac{r'}{b}\right)^{\alpha_\ell-1} + \left(\frac{b}{r'}\right)^{\alpha_\ell+1}\right] - \frac{C}{r'}\alpha_\ell\left(\frac{r'}{b}\right)^{\alpha_\ell-1}\left[\left(\frac{r'}{b}\right)^{\alpha_\ell} - \left(\frac{b}{r'}\right)^{\alpha_\ell}\right] = -\frac{1}{r'} \quad (A16)$$

which gives

$$C = -\frac{1}{2\alpha_\ell \varepsilon_{perp}}\frac{r'}{b} \quad (A17)$$

Therefore, for $\ell \neq 0$,

$$g_\ell(r,r') = -\frac{1}{2\alpha_\ell \varepsilon_{perp}}\left(\frac{r_<}{b}\right)^{\alpha_\ell}\left[\left(\frac{r_>}{b}\right)^{\alpha_\ell} - \left(\frac{b}{r_>}\right)^{\alpha_\ell}\right]\frac{r'}{r_< r_>}.$$

For $\ell = 0$,

$$g_0(r,r') = (1/2)\left(\frac{1}{r} - \frac{1}{b}\right). \quad (A18)$$

Consider a uniformly charged sphere of radius *a* and total charge *q*. The charge density is given by

$$\rho(r') = \frac{q}{4\pi a^2}\delta(r'-a). \qquad (A19)$$

The potential is given by

$$V = \int d^3r' G(r,r')\rho(r') = \frac{q}{4\pi\varepsilon_{perp}}\left(\frac{1}{r}-\frac{1}{b}\right), \qquad (A20)$$

our previous result. Since it is independent of *a*, it is also the result for a point charge at the origin. Now, let us consider the potential due to a point charge *q* at the point *z=d* on the z-axis. Then

$$\rho(r') = \frac{q}{2\pi d^2}\delta(r'-d)\delta(\cos\theta'-1), \qquad (A21)$$

which gives for the potential

$$V = \int d^3r\, G(\vec{r},\vec{r}')\rho(r') = \frac{q}{2\pi\varepsilon_{perp}}\left[(1/2)\left(\frac{1}{r}-\frac{1}{b}\right) - \frac{1}{8\pi r}\sum_{\ell=1}^{\infty}[2\ell+1]P_\ell(\cos\theta)\frac{1}{\alpha_\ell}\left(\frac{d}{b}\right)^{\alpha_\ell}\left[\left(\frac{r}{b}\right)^{\alpha_\ell}-\left(\frac{b}{r}\right)^{\alpha_\ell}\right]\right]$$

(A22).

The image potential energy is then

$$U_{image} = -\frac{q^2}{4\pi\varepsilon_{perp}}\left[\frac{1}{b}+\frac{1}{4\pi d}\sum_{\ell=1}^{\infty}\ell(\ell+1)\frac{1}{\alpha_\ell}\left(\frac{d}{b}\right)^{2\alpha_\ell}\right], \qquad (A23)$$

assuming that the $(b/d)^{\alpha_\ell}$ term in the last square bracket is part of the self-energy, since it gives a divergent contribution as $b/d \to \infty$. Also, the solution without a boundary only includes the $(b/d)^{\alpha_\ell}$ term. Then, $U_{image}$ can also be written as

$$U_{image} = -\frac{q^2}{4\pi\varepsilon_{perp}}\left[\frac{1}{b}+\frac{1}{4\pi d}\left(\frac{\varepsilon_{perp}}{\varepsilon_\parallel}\right)^{1/2}\sum_{\ell=1}^{\infty}[\ell(\ell+1)]^{1/2}v^{[\ell(\ell+1)]^{1/2}}\right], \qquad (A24)$$

where $v = (d/b)^{(\varepsilon_\parallel/\varepsilon_{perp})^{1/2}}$. Then,

$$U_{image} = -\frac{q^2}{4\pi\varepsilon_{perp}b}\left[1+A\left(\frac{\varepsilon_{perp}}{\varepsilon_\parallel}\right)^{1/2}\right], \qquad (A25)$$

where

$$A = \frac{b}{4\pi d} \sum_{\ell=1}^{\infty} [\ell(\ell+1)]^{1/2} v^{[\ell(\ell+1)]^{1/2}} . \quad (A26)$$

**Appendix B: Image Potential Energy for a Cylindrical Pore or Nanotube**

In order to determine the Coulomb potential in cylindrical coordinates for an ion lying on the cylinder's axis, we must solve[19]

$$\varepsilon_{perp}\left(\frac{\partial^2 V}{\partial \rho^2} + \frac{1}{\rho}\frac{\partial V}{\partial \rho}\right) + \varepsilon_{\|}\left(\frac{\partial^2 V}{\partial z^2}\right) = -\frac{q}{\rho}\delta(\rho)\delta(z). \quad (B1)$$

Making the substitution $\bar{\rho} = \rho/\varepsilon_{perp}^{1/2}$, $\bar{z} = z/\varepsilon_{\|}^{1/2}$, this equation becomes

$$\left(\frac{\partial^2 V}{\partial \bar{\rho}^2} + \frac{1}{\bar{\rho}}\frac{\partial V}{\partial \bar{\rho}}\right) + \left(\frac{\partial^2 V}{\partial \bar{z}^2}\right) = -\frac{1}{\varepsilon_{\|}^{1/2}\varepsilon_{perp}}\frac{q}{\bar{\rho}}\delta(\bar{\rho})\delta(\bar{z}). \quad (B2)$$

Since the solution to this equation without the factor $1/(\varepsilon_{\|}^{1/2}\varepsilon_{perp})$ is the Coulomb potential, the solution with this factor is

$$\frac{q}{4\pi\varepsilon_{\|}^{1/2}\varepsilon_{perp}(\bar{\rho}^2 + \bar{z}^2)^{1/2}}. \quad (B3)$$

Then, the solution can be written as

$$\frac{q}{4\pi\varepsilon_{perp}[\rho^2(\varepsilon_{\|}/\varepsilon_{perp}) + z^2]^{1/2}}. \quad (B4)$$

The solution by separation of variables $V = R(\rho)Z(z)$ is as follows:

$$\frac{1}{R(\rho)}\left(\frac{d^2 R(\rho)}{d\rho^2} + \frac{1}{\rho}\frac{dR(\rho)}{d\rho}\right) + \frac{\varepsilon_{\|}}{\varepsilon_{perp}}\frac{1}{Z(z)}\frac{d^2 Z(z)}{dz^2} = 0. \quad (B5)$$

Requiring that $Z(z)$ satisfy

$$\frac{d^2 Z(z)}{dz^2} = -k^2 Z(z), \quad (B6)$$

where $k^2$ is a constant, we find that $R(\rho)$ satisfies

$$\frac{d^2 R(\rho)}{d\rho^2} + \frac{1}{\rho}\frac{dR(\rho)}{d\rho} - \frac{\varepsilon_{\|}}{\varepsilon_{perp}}k^2 R(\rho) = 0, \quad (B7)$$

which is the equation for the modified Bessel function $I_0(uk\rho)$, where $u = (\varepsilon_{\|}/\varepsilon_{perp})^{1/2}$. Then, we can write

$$V(\rho, z) = \frac{q}{4\pi\varepsilon_{perp}[(\varepsilon_{\|}/\varepsilon_{perp})\rho^2 + z^2]^{1/2}} + (1/2)\int_{-\infty}^{\infty} dk A(k) e^{ikz} I_0(uk\rho). \quad (B8)$$

We require that

$$0 = V(b, z) = \frac{q}{4\pi\varepsilon_{perp}[(\varepsilon_{\|}/\varepsilon_{perp})b^2 + z^2]^{1/2}} + (1/2)\int_{-\infty}^{\infty} dk A(k) e^{ikz} I_0(ukb), \quad (B9)$$

where $b$ is the radius of the pore. Then,

$$A(k) I_0(ukb) = -\frac{2q}{4\pi\varepsilon_{perp}} \frac{1}{2\pi} \int_{-\infty}^{\infty} dz \frac{e^{-ikz}}{[(\varepsilon_{\|}/\varepsilon_{perp})b^2 + z^2]^{1/2}}. \quad (B10)$$

Therefore, the image potential energy is given by

$$U_{image} = -\frac{q^2}{8\pi\varepsilon_{perp}} \frac{4}{\pi} \int_0^\infty \frac{dk}{I_0(kub)} \int_0^\infty dz' \frac{\cos(kz')}{[u^2 b^2 + z'^2]^{1/2}}, \quad (B11)$$

or setting $k'=kb$ and $z''=z'/b$, we get

$$U_{image} = -\frac{q^2}{8\pi\varepsilon_{perp} b} \frac{4}{\pi} \int_0^\infty \frac{dk'}{I_0(k'u)} \int_0^\infty dz'' \frac{\cos(k'z'')}{[u^2 + z''^2]^{1/2}}. \quad (B12)$$

Let us now consider a point charge which is a distance $d$ off the axis (i.e., it is located at $x=d$). The potential due to the point charge satisfies

$$\varepsilon_{perp}\left(\frac{\partial^2 V}{\partial \rho^2} + \frac{1}{\rho}\frac{\partial V}{\partial \rho}\right) + \varepsilon_{\|}\left(\frac{\partial^2 V}{\partial z^2} + \frac{1}{\rho^2}\frac{\partial^2 V}{\partial \phi^2}\right) = -\frac{q}{\rho}\delta(\rho - d)\delta(\phi)\delta(z), \quad (B13)$$

or

$$\left(\frac{\partial^2 V}{\partial \rho^2} + \frac{1}{\rho}\frac{\partial V}{\partial \rho}\right) + u\left(\frac{\partial^2 V}{\partial z^2} + \frac{1}{\rho^2}\frac{\partial^2 V}{\partial \phi^2}\right) = -\frac{q}{\varepsilon_{perp}\rho}\delta(\rho - d)\delta(\phi)\delta(z) \quad (B14)$$

where $u = \varepsilon_{\|}/\varepsilon_{perp}$. The solution by separation of variables $V = R(\rho)Z(z)Q(\phi)$ is as follows:

$$\frac{1}{R(\rho)}\left(\frac{d^2 R(\rho)}{d\rho^2} + \frac{1}{\rho}\frac{dR(\rho)}{d\rho}\right) + u\left(\frac{1}{Z(z)}\frac{d^2 Z(z)}{dz^2} + \frac{1}{\rho^2}\frac{\partial^2 Q}{\partial \phi^2}\right) = 0. \quad (B15)$$

Requiring that $Z(z)$ satisfy

$$\frac{d^2 Z(z)}{dz^2} = -k^2 Z(z), \quad (B16)$$

where $k^2$ is a constant, and

$$\frac{d^2Q}{d\phi^2} = -m^2 Q, \qquad (B17)$$

we find that $R(\rho)$ satisfies

$$\frac{d^2 R(\rho)}{d\rho^2} + \frac{1}{\rho}\frac{dR(\rho)}{d\rho} - u\left(k^2 + \frac{m^2}{\rho^2}\right)R(\rho) = 0, \qquad (B18)$$

whose solutions are $I_{mu^{1/2}}(u^{1/2}k)$, $K_{mu^{1/2}}(u^{1/2}k)$. Then, the Green's function can be written as

$$G = \frac{1}{2\pi^2}\sum_{m=-\infty}^{\infty}\int_0^{\infty} dk\, e^{im(\phi-\phi')}\cos k(z-z') g_m(\rho,\rho'), \qquad (B19)$$

where $g_m(\rho,\rho')$ satisfies

$$\frac{1}{\rho}\frac{d}{d\rho}\left(\rho\frac{dg_m}{d\rho}\right) - u\left(k^2 + \frac{m^2}{\rho^2}\right)g_m = -\frac{1}{\rho}\delta(\rho-\rho'). \qquad (B20)$$

Integrating from $\rho = \rho' - \delta$ to $\rho = \rho' + \delta$, where $\delta \ll 1$, we get

$$\frac{dg_m}{d\rho}\bigg|_{\rho=\rho'+\delta} - \frac{dg_m}{d\rho}\bigg|_{\rho=\rho'-\delta} = -\frac{1}{\rho'}. \qquad (B21)$$

The solution is

$$g_m = C I_{mu^{1/2}}(x_<) K_{mu^{1/2}}(x_>), \qquad (B22)$$

where $C$ is a constant, $x = ku^{1/2}\rho$, $\rho_<$ is the smaller of $\rho, \rho'$ and $\rho_>$ is the larger of $\rho, \rho'$. Then, we have

$$C\left[I_{mu^{1/2}}(x')\frac{dK_{mu^{1/2}}(x)}{dx} - K_{mu^{1/2}}(x)\frac{dI_{mu^{1/2}}(x)}{dx}\right]_{x=x'} = -\frac{1}{x'}. \qquad (B23)$$

Since the bracketed expression (the Wronskian) is a constant for the equation satisfied by $g_m$, we can evaluate it at one point, such as the in the large $\rho$ limit, which gives a value of $-x'^{-1}$, and hence, we find that $C=1$. Thus, from the above expression for the Green's function we find that the potential due to a point charge located a distance $d$ from the cylinder axis along the x-axis (i.e., $\rho' = d, \phi' = 0, z' = 0$) is given by

$$V = \frac{q}{2\pi^2 \varepsilon_{perp}}\sum_{m=-\infty}^{\infty} e^{im\phi}\int_0^{\infty} dk \cos kz \left[I_{mu^{1/2}}(ku^{1/2}\rho_<)K_{mu^{1/2}}(ku^{1/2}\rho_>) + A_m(k)I_{mu^{1/2}}(ku^{1/2}\rho)\right]. \qquad (B24)$$

The coefficient $A_m(k)$ is obtained by requiring that $V(\rho = b, z, \phi) = 0$, which gives

$$A_m(k) = -\frac{I_{mu^{1/2}}(ku^{1/2}d) K_{mu^{1/2}}(ku^{1/2}b)}{I_{mu^{1/2}}(ku^{1/2}b)}.  \qquad (B25)$$

Inserting this expression for this coefficient and substituting $\rho = d$ and multiplying the second term in the square bracket by q we obtain for the image potential

$$U_{image} = -\frac{q^2}{8\pi\varepsilon_{perp}b}\frac{4}{\pi}\sum_{m=0}^{\infty}\int_0^{\infty}dk'\frac{I_{mu^{1/2}}(k'u^{1/2}d/b)^2 K_{mu^{1/2}}(k'u^{1/2})}{I_{mu^{1/2}}(k'u^{1/2})}, \qquad (B26)$$

where *k'=kb*. In performing the integral over k', the upper limit is chosen as the value of k' at which the integrand becomes negligibly small. Similarly, the upper limit on the summation over m was chosen as the value of *m* at which the summand becomes negligibly small.